\documentstyle[amssymb,prb,aps,epsfig,multicol]{revtex}

\begin{document}
\title{Superconductivity and Electronic Structure of Perovskite MgCNi$_3$}
\author{D.J. Singh and I.I. Mazin}
\address{Center for Computational Materials Science,\\
Naval Research Laboratory, Washington, DC 20375, U.S.A.}
\date{\today}
\maketitle

\begin{abstract}
The electronic structure,
stability, electron phonon coupling and superconductivity of the non-oxide
perovskite MgCNi$_3$ are studied using density functional calculations.
The band structure is dominated by
a Ni $d$ derived density of states peak just below the Fermi energy,
which leads to a moderate Stoner enhancement, placing MgCNi$_3$ in the
range where spin fluctuations may noticeably affect transport, specific
heat and superconductivity, providing a mechanism for reconciling various
measures of the coupling $\lambda$. Strong electron phonon interactions
are found for the octahedral rotation mode and may exist for
other bond angle bending modes.
The Fermi surface contains nearly cancelling hole and electron sheets
that give unusual behavior of transport quantities particularly the
thermopower.
The results are discussed in relation
to the superconductivity of MgCNi$_3$.
\end{abstract}

\begin{multicols}{2}


The discovery \cite{he} of superconductivity
in the non-oxide perovskite MgCNi$_{3}$,
with critical temperature $T_{c}\approx $8K, raises the questions of
how superconductivity appears in such a Ni rich phase and its
relationship to other unusual superconducting phases, particularly the
Ni, Pd and Pt borocarbides and boronitrides \cite%
{nagarajan,cava1} and PdH. %
\cite{skosk,papa,rowe,rahman,klein} These possibly related materials have
relatively high values of $T_{c}$ and high concentrations of magnetic or
near magnetic group 8 elements but different underlying physics.

Typically, but not always, perovskites are distorted by the freezing
in of unstable zone boundary rotational phonons to form structures like the P%
$nma$, GdFeO$_3$ structure or off-centerings like those that produce
ferroelectricity as in BaTiO$_3$. This is due to
the importance of ionic and central interactions in many perovskites;
the stability of the ideal cubic structure depends on a balance between 
$A$-site cation--anion and the $B$-site cation--anion interactions.
In the minority that are truely cubic,
there are often soft anharmonic phonons of displacive character,
as $e.g.$ in KTaO$_3$ and/or rotational character (the combination common in
Pb based perovskites). \cite{fornari}

According to refinements, \cite{he} MgCNi$_{3}$ is stoichiometric
with only a very small, 4\% carbon deficiency and occurs in the ideal
cubic perovskite structure. The ambient temperature lattice parameter is
3.812\AA , which yields a C-Ni bond length of 1.906\AA . Considering
that Mg is expected to occur as Mg$^{2+}$ and so the CNi$_{3}$ subunit as
negatively charged, this bond may be considered short. As such,
strong Ni--C covalent interactions are expected. If so and if
bonding or antibonding electronic states associated with this
are present at the Fermi energy $E_{F}$, strong electron phonon coupling
(EPC) may be expected.
Theoretical studies of the electronic structure and phonons showed that in
fact EPCs associated with relatively hard modes that modulate strong bonds
involving Ni and first row elements play an important role in the
superconductivity of the borocarbides and nitrides. \cite%
{mattheiss,pickett,singh}

On the other hand, the perovskite topology and the fact that
Mg likely acts as a simple cation that is electronically inactive near $%
E_{F}$ suggests that perhaps low frequency anharmonic modes play an
important role. In particular, phonon modes, like that associated with the
octahedral rotation, will modulate the C-Ni-C bond angles, which in this
structure are expected to be important in the hopping associated with
metallic conduction.
The crystal structure may be conceptually viewed as expanded $fcc$
Ni with 25\% of the sites replaced by Mg and C interstitials in 
octahedral sites. The bands thus should be 
narrow and transition metal-%
like around $E_F$ with a higher filling than in pure Ni.
This suggests some analogy with the band structure of
PdH\cite{papa,klein}, also an expanded group 8 metal with
a higher filling. The main difference is the importnat role 
of C as hopping mediator in  MgCNi$_{3}$, as discussed
below.

The present experimental situation is confusing. Tunneling and
upper critical field measurements were reported by Mao 
{\it et al.}\cite{mao} These characterize MgCNi$_{3}$ as a strong coupling
superconductor with an unusually large reduced energy gap, $2\Delta
(0)/kT_{c}\approx 10,$ near the upper theoretical
limit \cite{carbotte} for $s$-wave
superconductivity. However, the specific heat jump $\Delta
C(T_{c})/kT_{c}\approx 1.9$ suggests moderate $\lambda \lesssim $1 coupling. %
\cite{carbotte} Furthermore, they suggested that MgNiC$_{3}$ may be a non-$s$
superconductor, based on observation of a zero-bias anomaly (often,
but not always, due to a sign-changing order parameter). This
implies an analogy with metallic Pd where superconductivity is not observed,
but a possible spin fluctuation mediated mechanism was discussed, or Sr$%
_{2}$RuO$_{4}$, which is apparently a triplet superconductor due to spin
fluctuations. \cite{maeno} On the other hand, doping with Co, a likely
magnetic impurity, rapidly decreases the superconducting volume, \cite%
{hayward,ren} and the same effect, but stronger, is found with Mn. \cite%
{ren} Cu doping, however, reduces $T_{c}$ without suppressing the
superconducting fraction.\cite{hayward} This is consistent with
expectations for $s$-wave pairing in a rigid band model. Hayward {\it et al.}%
\cite{hayward} reported a calculated electronic
density of states (DOS)
that shows a large peak just below $E_{F}$ and speculated
about the importance of magnetic fluctuations based on this. \cite{hayward}
In this regard, doping on the Mg $A$-site may be illuminating as rigid band
behavior may be more likely in that case.

Here we report electronic structure calculations focusing on
these issues. These were done in the local density
approximation (LDA) \cite{hl} with the general potential linearized
augmented planewave (LAPW) method. \cite{singh-book} Well converged
zone samplings and basis sets, including local orbitals \cite%
{singh-lo} to treat the Mg semicore states and relax linearization errors
for the Ni $d$ bands were used. The experimental lattice constant of $a$%
=3.82\AA~was used (our calculated LDA lattice parameter is 1.7\% smaller as
is typical for this approximation). We study
the band structure and Fermiology, the susceptibility and proximity to
magnetism, transport coefficients and selected phonon modes and
electron-phonon couplings. Transport coefficients were determined using
kinetic theory \cite{ziman} with zone integrations using
LAPW eigenvalues on a grid of 2925 {\bf k}-points in the
irreducible $1/48$ wedge.

\begin{figure}[tbp]
\centerline{\epsfig{file=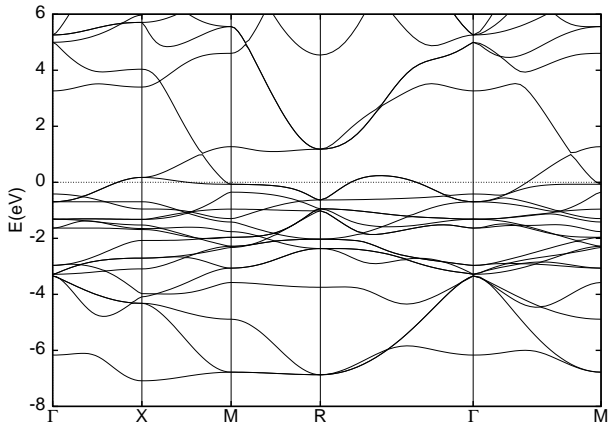,angle=270,width=0.92\linewidth}}
\vspace{0.125in}
\setlength{\columnwidth}{3.2in} \nopagebreak
\caption{LDA band structure of MgCNi$_3$.
$E_F$ is at 0.}
\label{bands}
\end{figure}

\begin{figure}[tbp]
\centerline{\epsfig{file=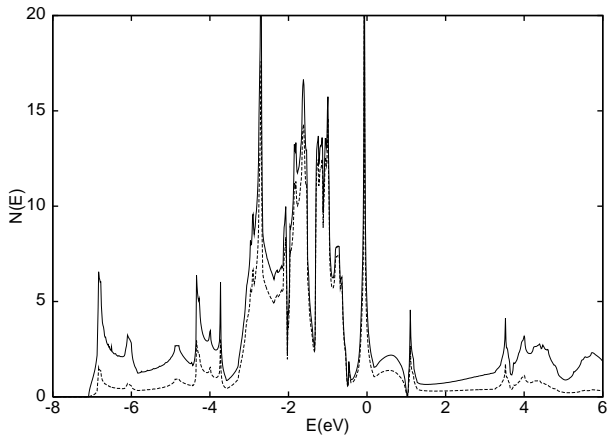,angle=270,width=0.92\linewidth}}
\vspace{0.125in}
\centerline{\epsfig{file=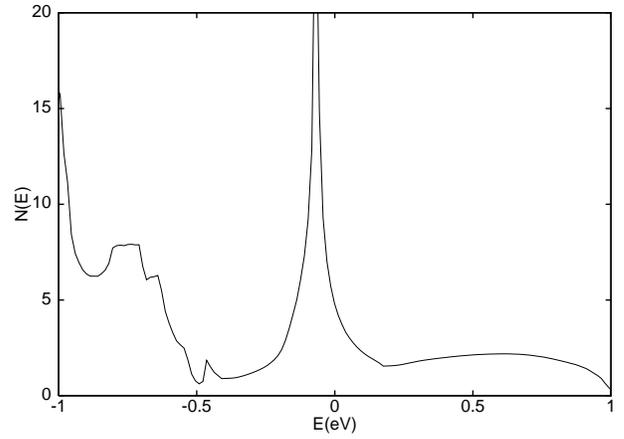,angle=270,width=0.92\linewidth}}
\vspace{0.125in}
\setlength{\columnwidth}{3.2in} \nopagebreak
\caption{Electronic DOS (upper panel) of MgCNi$_3$ as calculated
within the LDA.
The dashed line is the $d$ contribution within the Ni LAPW spheres of
radius 2.04 a$_0$. The lower panel is a blow-up around $E_F$.}
\label{dos}
\end{figure}

\begin{figure}[tbp]
\centerline{\epsfig{file=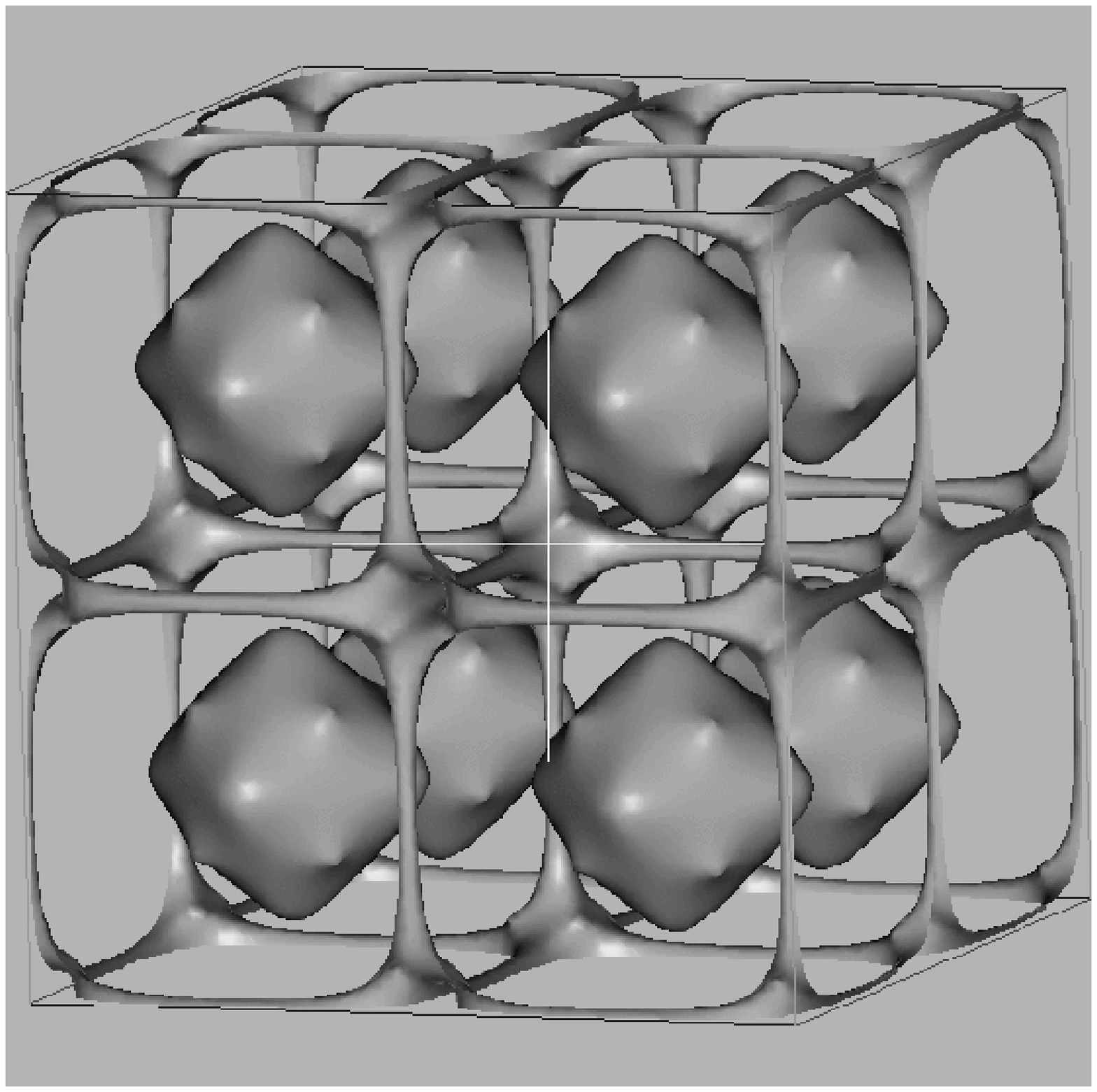,angle=0,width=0.92\linewidth}}
\vspace{0.125in}
\centerline{\epsfig{file=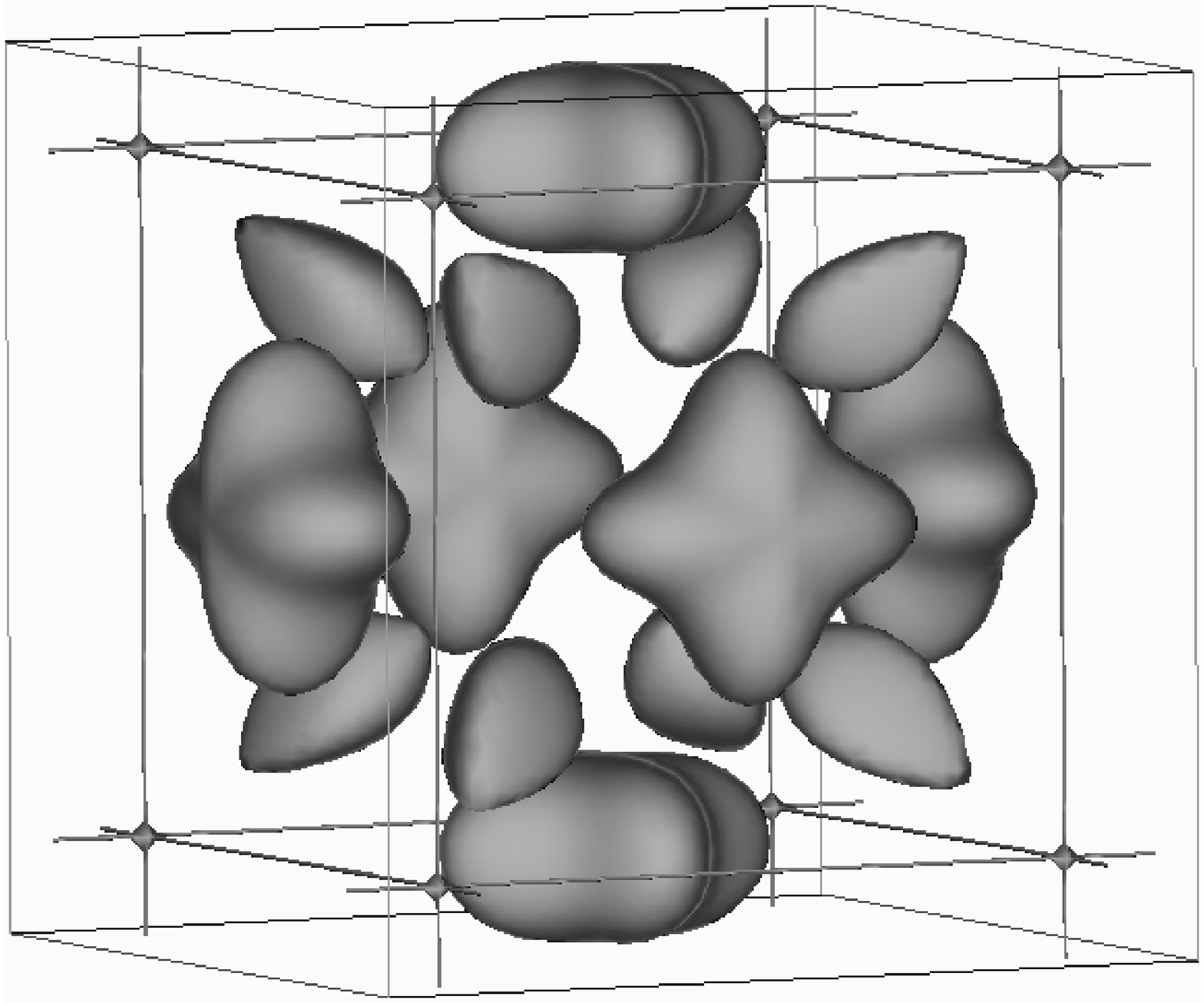,angle=0,width=0.92\linewidth}}
\vspace{0.125in}
\setlength{\columnwidth}{3.2in} \nopagebreak
\caption{Calculated Fermi surfaces.
The top panel is from the lower band and shows eight zones.
The rounded cube sections are centered at $\Gamma$, while
the thin jungle gym is along the $M-R$ lines.
The bottom panel is from the upper band, showing dimpled square shaped sections
centered at $X$ and ovoids along $\Gamma-R$.}
\label{fermi}
\end{figure}

The band structure and DOS are
shown in Figs. \ref{bands} and \ref{dos}. The Fermi surfaces are in
Fig. \ref{fermi}. As expected, Mg plays an minor role in the bands in the
valence region. (There is a weak Ni $d$-Mg $s$ bonding interaction.
The $\Gamma$ point state at -6 eV has Mg-Ni bonding character with the
corresponding antibonding state at +3 eV. This covalency is localized around 
$\Gamma$).
Octahedral coordination is unfavorable for formation of C $sp$ hybrids,
so not surprisingly the C $s$ orbitals are also
inactive. They occur around -12 eV (relative to $E_F$) and are not shown.
The valence region, which extends from -7.1 eV to +1 eV is
therefore derived from the 15 Ni $d$ and 3 C $p$ orbitals filled by 34
electrons. The C $p$ orbitals are strongly hybridized with Ni $d$, and are
located below most of the $d$ states. So in the first approximation they can
be integrated out and the bands near the Fermi level can be analyzed in
terms of the Ni $d$ states.

The unusual two-fold linear Ni coordination by C makes some of the bands
very narrow. For instance, Ni(x) $yz$ and $y^{2}-z^{2},$ Ni(y) $zx$ and $%
z^{2}-x^{2},$ and Ni(x) $xy$ and $x^{2}-y^{2}$ orbitals do not disperse
in the nearest neighbor approximation (NNA) as they have no C orbitals to hop
through. \cite{coord-note} The remaining 9 orbitals form three independent
manifolds, consisting of Ni(x) $xy$, Ni(z) $yz$, and Ni(y) $y^{2}-r^{2}$
orbitals, coupled via C $p_{y}$ and the two corresponding combinations ({\it %
i.e.} $p_{x}$ and $p_{z}$). Interestingly, in the NNA,
one of the three resulting bands in each manifold, the one
that involves the antibonding combination of the $t_{2g}$ orbitals, is
non-bonding again. The result is 3 bonding, 3 nonbonding and 3 antibonding
bands. Two of these antibonding bands cross $E_{F}$. In the large crystal
field limit (relative to the small widths), these have a simple dispersion
proportional to $\sin (ak_{x}/2)^{2}+\sin (ak_{y}/2)^{2}$ etc., which
naturally makes them flat for a number of high-symmetry directions.
This is reflected in the Fermi surface (Fig. \ref{fermi}). The lower band
produces $\Gamma $ centered rounded cube shaped electron sections and a
narrow low weight jungle gym along the $R-M$ lines also containing
electrons. The upper band (Fig. \ref{fermi} bottom) forms a two-sheet Fermi
surface consisting of pancaked squares centered at the $X$ points and
ovoids along the $\Gamma -R$ lines, both hole-like. The flat square shape
reflects weak dispersion of the underlying band along $X-M$ and strong
dispersion along $X-\Gamma $ (The quasi-2D behavior is because $\sin
(ak_{x}/2)^{2}+\sin (ak_{y}/2)^{2}$ does not disperse along $z$.) Near $%
\Gamma $, where these bands make up the three fold degenerate state (this is
the second one below $E_{F}$), the dispersion is due to direct Ni-Ni $%
dd\sigma $ hopping. Turning to the two bands that cross $E_{F}$, the lower
band forms the more interesting heavy sheet of Fermi surface around $X$. At
an $M$ point, say (110), $\sin (ak_{x}/2)^{2}+\sin (ak_{z}/2)^{2}$ and $\sin
(ak_{y}/2)^{2}+\sin (ak_{z}/2)^{2}$ are degenerate and their
hybridization via Ni(x) $xy$ and Ni(y) $xy$ orbitals is proportional to $%
\cos (ak_{x}/2)\cos (ak_{y}/2)$ and vanishes near $M,$ which is why they
are heavy. On lowering $E_F$, the $X$ centered squares grow
grow, as does DOS at $E_F$, $N(E_F)$,
until they meet at the $M$ points and the topology
changes.

The value of $N(E_{F})$ is 4.99 eV$^{-1}$ on a per
formula unit basis. Two recent non-full-potential studies obtain differing
values. Dugdale and Jarlborg report 6.35 eV$^{-1}$ and 3.49 eV$^{-1}$
depending on the exact method they employ, \cite{dugdale} while Shim and Min
obtain 5.34 eV$^{-1}$, which is much closer to the present result. \cite%
{shim} These differences are significant as they control the proximity to
magnetism. In particular, the susceptibility, $\chi (0)$, is determined by $%
N(E_{F})/(1-N(E_{F})I)$, where $I$ is the Stoner parameter, and the
denominator contains a small difference involving $N(E_{F})$. Comparing our
calculated full potential $N(E_{F})$ with the experimental linear
coefficient $\gamma \approx $10 mJ/moleNiK$^{2}$, we obtain a specific heat
renormalization $\gamma /\gamma _{band}$=2.6. The calculated plasma
frequency is $\hbar \omega _{P}$ = 3.25 eV. The calculated Hall number is
-1.3$\times $10$^{22}$ cm$^{-3}$. This agrees well with the measured value
of Li {\it et al.}\cite{li} The constant scattering time approximation
thermopower, $S$, is $p$-type except at very low $T$ below 10K where it is $n
$-type consistent with the Hall number. $S$ is very small (less than 1 $\mu $%
V/K) below 150K but then rises more rapidly reaching 5 $\mu $V/K at 300K and
16 $\mu $V/K at 600K. Both the Hall number and the thermopower
are controlled by
the competition between the hole and electron pockets of comparable size,
which leads, for instance, to the unusual $T-$dependence of thermopower.
Using the calculated $\omega _{P}$ and the measured resistivity %
\cite{he,li} in the Bloch-Gruneisen formula, $d\rho /dT$=$(8\pi ^{2}/\hbar
\omega _{P}^{2})k_{B}\lambda _{tr}$, we obtain $\lambda _{tr}\approx 1-1.6$,
though we note that this extraction of $\lambda _{tr}$
is sensitive to sample quality.

The simplest estimate of $\lambda$ from zone center frozen-phonon
calculations is not applicable to cubic perovskites since no
zone-center modes couple by symmetry. However
one can use frozen-phonon zone-corner calculations for this purpose, as in
Ref.\onlinecite{BKB}. The relevant formula is

\begin{eqnarray}
\lambda=\sum_{\nu{\bf q}}\lambda_{\nu{\bf q}} &\approx& \frac{2}{%
N_{\uparrow}\omega_{\nu{\bf q}}} \sum_{\nu}\langle |g^\nu|^2 \rangle
\sum_{n,m,{\bf k,q}}\delta(\varepsilon_{n,{\bf k+q}}) \delta(\varepsilon_{m,%
{\bf k}})  \nonumber \\
&=&2 N_{\uparrow}\sum_{\nu}\langle |g^\nu|^2 /\omega_{\nu{\bf q}} \rangle 
\nonumber
\end{eqnarray}

Here $N_{\uparrow }$ is the per spin DOS at $E_{F}$, and $g^{\nu }$ is an
electron-ion matrix element from the derivative of the ionic potential with
respect to the dimensionless phonon coordinate (see Ref. \onlinecite{BKB}
for details). To get qualitative information on the EPC, we focus
on two R-point phonons. These are the octahedral rotation and the fully
symmetric breathing mode. The former mode changes the C-Ni-C bond angles,
but not the bond lengths in lowest order, while the latter is a pure bond
stretching mode. The calculated frequencies are 105 cm$^{-1}$ and 349 cm$%
^{-1}$, respectively.

To estimate the matrix element, we selected several points on the
intersection line of the {\bf k} and {\bf k+q} Fermi surfaces, and fitted
the bands in the nearby region with the second-order perturbation theory,
and then averaged the resulting matrix element. We included three
intersection points, where $g$ does not vanish by symmetry, for the
breathing mode, and 7 points for the rotational mode. Assuming that
averaging over ${\bf q}$ does not change $g^{2}$, we find for the breathing
mode $\lambda \approx 0.005$ and for the rotational mode $\lambda \approx 1.2
$. The latter dominates not only because of its nearly 4 times larger
deformation potential and its 15 times smaller denominator $M\omega ^{2}$.
Because of the mode degeneracy, this corresponds to a total rotational modes 
$\lambda _{rot}\approx 3.6$ However, in oxide perovskites, this $R_{25}$
rotation mode usually stiffens rapidly away from the zone boundary,
reflecting the rigidity of the O octahedra. Here such a stiffening may also
be expected. Thus, the zone averaged $\lambda $ will
likely be considerable smaller, but probably still substantial. In any case,
it can be said that the stiff C-Ni bond stretching modes are apparently not
significant contributors to the EPC, while rotational (and probably other
C-Ni-C bond bending) modes are strongly coupled. Note that the popular rigid
muffin-tin potential method is hard to apply here: for transition metals, it
is known to overestimate EPC, and Ni is in a low-symmetry position. Most
importantly, with such a large disparity of the contributions from different
modes, there is no telling beforehand, which average phonon frequency should
be used in calculating $\lambda .$ 

As mentioned, possible nearness a magnetism
could conceivably play
an important role. To investigate this, we have calculated spin
susceptibility directly from the variation of the total energy with small
imposed magnetizations (ranging from 0.1 to 0.7 $\mu _{B}$ per cell) in
fixed spin moment
calculations, $\delta E_{tot}=\chi
^{-1}m^{2}/2=(N_{\uparrow }^{-1}-I)m^{2}/4,$ where $I$ is the Stoner factor,
characterizing intraatomic exchange (for a compound like MgCNi$_{3},$ $I$ is
expected to be close to 1/3 of the pure Ni value of 1.16 eV, probably
slightly reduced because of hybridization with C).
It appears that $\delta
E_{tot}(m)$ noticeably deviates from the quadratic behavior. This is
expected in extended Stoner theory,
  \cite{krasko}
where an
average $\tilde{N}(m)$ is substituted for $N_{\uparrow }.$ With  $\tilde{N}%
(m),$ the above expression gives a good
description of  $\delta E_{tot}(m)
$ up to $m\leq 0.6$ $\mu _{B}$ with a
weakly energy dependent $I\approx 0.95/3$ eV.
Self-consistent virtual crystal calculations corresponding to a 10\%
replacement of Mg by a monovalent element (Na in the calculations but
with the lattice parameter fixed - so Li or 
a 5\% Mg deficiency may be a better experimental
analogue) produced a borderline ferromagnetic ground state with a very low
energy gain and a moment of 0.25 $\mu _{B}$/cell.

The susceptibility renormalization $(1-IN_{\uparrow
})^{-1}\approx 5,$ and $\chi =2.7\times10^{-4}$ emu/moleNi,
somewhat larger
than the reported experimental value\cite{hayward} of 1.7/times10$^{-4}$
emu/moleNi. This number is expected to be very sensitive to doping,
{\it i.e.} composition, due to the near cancelation
in the denominator. Perhaps
the difference with experiment can be understood in these terms {\it e.g.}
if the 4\% of C vacancies lead to a higher effective band filling.
If the Stoner renormalization is indeed $\approx 5,$ this
signals presence of significant spin fluctuations. To compare with, in 
Sr$_{2}$RuO$_{4}$,
where $(1-IN_{\uparrow })^{-1}\approx 9,$ spin fluctuations are
believed to cause triplet superconductivity, and in Pd metal, where it is $%
\approx 30,$ spin fluctuations destroy superconductivity which would
otherwise exist due to the sizeable ($\lambda \gtrsim 0.5)$ EPC.
Importantly, the effect of spin fluctuations on mass renormalization,
transport and superconducting properties is very different: essentially, in
the first two cases, the coupling constants add, while for
superconductivity they add in the mass renormalization term, and subtract in
the pairing term (In the strong coupling $T_{c}$ is not proportional
to $\exp [-(1+\lambda )/\lambda ],$ but rather $\exp [-(1+\lambda
_{ph}+\lambda _{spin})/(\lambda _{ph}-\lambda _{spin})]$.) This may explain
the inconsistency between coupling constants determined from
different experiments.

In any case, the present results underscore certain similarities with
PdH in terms of band structure and magnetic renormalizations,
as well as with the borocarbides. However, the fact 
that the most important phonons are Ni bond-bending modes 
places MgCNi$_3$ in a unique class of its own.


This work is supported by ONR and the ASC computer center.

\end{multicols}


\begin{references}

\bibitem{he} T. He,
{\it et al.},
Nature {\bf 411}, 54 (2001).

\bibitem{nagarajan} R. Nagarajan,
{\it et al.},
Phys. Rev. Lett. {\bf 72}, 274 (1994).

\bibitem{cava1} R.J. Cava,
{\it et al.},
Nature {\bf 367}, 146 (1994);
R.J. Cava, {\it et al.},
Nature {\bf 367}, 252 (1994).
R.J. Cava, {\it et al.},
Phys. Rev.  B {\bf 49}, 12384 (1994);
R.J. Cava, {\it et al.},
Nature {\bf 372}, 245 (1994).

\bibitem{skosk} T. Skoskiewicz, Phys. Status Solidi A {\bf 11}, K123 (1972).
\bibitem{papa} D.A. Papaconstantopoulos, B.M. Klein, J.S. Faulkner, and L.L.
Boyer, Phys. Rev. B {\bf 18}, 2784 (1978).

\bibitem{rowe} J.M. Rowe,
{\it et al.},
Phys. Rev. Lett. {\bf 33}, 1297 (1974).

\bibitem{rahman} A. Rahman,
{\it et al.},
Phys. Rev. B {\bf 14}, 3630 (1976).

\bibitem{klein} B.M. Klein, and R.E. Cohen, Phys. Rev. B {\bf 45}, 12405
(1992).



\bibitem{fornari} M. Fornari, and D.J. Singh, Phys. Rev. B {\bf 63}, 092101
(2001).

\bibitem{mattheiss} L.F. Mattheiss, T. Siegrist, and R.J. Cava, Solid State
Commun. {\bf 91}, 587 (1994).

\bibitem{pickett} W.E. Pickett, and D.J. Singh, Phys. Rev. Lett. {\bf 72},
3702 (1994).

\bibitem{singh} D.J. Singh, and W.E. Pickett, Phys. Rev. B {\bf 51}, 8668
(1995).

\bibitem{mao} Z.Q. Mao,
{\it et al.},
cond-mat/0105280.

\bibitem{carbotte} J.P. Carbotte, Rev. Mod. Phys. {\bf 62}, 1027 (1990).

\bibitem{maeno} Y. Maeno,
{\it et al.},
Nature {\bf 372}, 532 (1994).

\bibitem{hayward} M.A. Hayward,
{\it et al.},
cond-mat/0104541.

\bibitem{ren} Z.A. Ren,
{\it et al.},
cond-mat/0105366.

\bibitem{hl} L. Hedin, and B. I. Lundqvist, J. Phys. C {\bf 4}, 2064 (1971).

\bibitem{singh-book} D.J. Singh, {\it Planewaves, Pseudopotentials and the
LAPW Method} (Kluwer Academic, Boston, 1994).

\bibitem{singh-lo} D. Singh, Phys. Rev. B {\bf 43}, 6388 (1991).

\bibitem{ziman} J.M. Ziman, Principles of the Theory of Solids (Cambridge
University Press, Cambridge, 1972), and references therein.

\bibitem{coord-note} Ni has tetragonal symmetry with the axis directed
towards the two neighboring C. We use Ni(x) to denote the Ni
with tetragonal axis along x.

\bibitem{dugdale} S.B. Dugdale, and T. Jarlborg, preprint (cont-mat/0105349).

\bibitem{shim} J.H. Shim, and B.I. Min, preprint (cont-mat/0105418).

\bibitem{li} S.Y. Li,
{\it et al.},
cond-mat/0104554.

\bibitem{BKB} A.I. Lichtenstein,
{\it et al.},
Phys. Rev. B {\bf 44}, 5388 (1991).


 \bibitem{krasko}
 G.L. Krasko, Phys. Rev. B {\bf 36}, 8565 (1987).

\end{references}
\end{document}